\documentclass[12pt]{article}
\usepackage[T1]{fontenc}
\usepackage{csquotes}
\usepackage{sbc-template}

\usepackage[dvipdfm]{graphicx}
\usepackage{url}
\usepackage{multicol}

\usepackage{hyperref}
\usepackage{svg}
\usepackage[utf8]{inputenc}  
\usepackage[editing]{coop-writing}
\makeatletter
\newcommand{\keywords}[2][Keywords.]{%
      {\list{}{
      \relax
      \leftmargin=0.8cm
      \labelwidth=\z@
      \listparindent=\z@
      \setlength\topsep{-20 pt}
      \itemindent\listparindent
      \rightmargin\leftmargin}%
      \item[\hskip\labelsep                                    \bfseries\itshape #1]\itshape#2%
      \endlist}}
\makeatother
\cwnamedef{xexeo}{red}{Xexéo}
\cwnamedef{marcus}{blue}{Marcus}
\cwdefanoncitetext{[0]}
\cwdefanontext{[AVALIAÇÃO CEGA]}
\hypersetup{colorlinks=false}
\hypersetup{pdfborder = {0 0 0}}
     
\title{A Vocabulary of Board Game Dynamics}

\author{ {Joshua Kritz\inst{1} }{Geraldo Xexéo\inst{1}} }

\address{
{LUDES - Programa de Engenharia de Sistemas e Computação} \\
{COPPE - Universidade Federal do Rio de Janeiro } \\ 
{Avenida Horácio Macedo, 2030, CT, Bloco H, sala 319, Rio de Janeiro, RJ - Brasil}
\email{{ joshuakritz2@gmail.com, xexeo@cos.ufrj.br}}
}
\begin{document} 

\maketitle

\begin{abstract}
  In recent years, significant advances have been made in the field of game research. However, there has been a noticeable dearth of scholarly research focused on the domain of dynamics, despite the widespread recognition among researchers of its existence and importance. The objective of this paper is to address this research gap by presenting a vocabulary dedicated to boardgame dynamics. To achieve this goal, we employ a focus group to generate a set of dynamic concepts that are subsequently subjected to validation and refinement through a survey. The resulting concepts are then organized into a vocabulary using a taxonomic structure, allowing the grouping of these concepts into broader and more general ideas.
\end{abstract}
     
\begin{resumo} 
  Nos últimos anos, avanços significativos foram feitos no campo da pesquisa de jogos. No entanto, tem havido uma notável escassez de investigações acadêmicas no domínio de dinâmicas, apesar do amplo reconhecimento entre os pesquisadores de sua existência e importância. O objetivo deste trabalho é cobrir essa lacuna de pesquisa, apresentando um vocabulário de dinâmica de jogos de tabuleiro. Para isso, empregamos um grupo focal para gerar um conjunto de conceitos, que são posteriormente submetidos a validação e refinamento por meio de um questionário. Os conceitos resultantes são então organizados em um vocabulário usando uma estrutura taxonômica, possibilitando o agrupamento desses conceitos em ideias mais amplas e gerais.
\end{resumo}

\keywords{Games, Game Dynamics, Vocabulary}    

\section{Introduction}

This work aims to address a long-standing gap in the field of game studies, specifically focusing on game dynamics. The concept of dynamics is understood in accordance with the MDA framework \cite{Hunicke2004}, which defines it as ``the runtime behavior of game mechanics'' \cite{Hunicke2004}. 
While there is a wealth of research on game mechanics, evidenced by the extensive body of work available, such as Schell's "The Art of Game Design" \cite{schell2014art}, Salen and Zimmerman's "Rules of Play" \cite{salen2004rules}, and Koster's "The Theory of Fun" \cite{koster2013theory}, among others, these works primarily focus on explaining and developing other aspects of games. 
Consequently, they do not provide a clear definition of dynamics or a substantial collection of examples. 
Consequently, there is a lack of resources for accessing a comprehensive set of terms and concepts pertaining to game dynamics, in contrast to the availability of such resources for mechanics and aesthetics.

Dynamics play a pivotal role in games as they establish the connection between the mechanics devised by game designers and the ultimate experience of players when engaging with the game.
Therefore, understanding how dynamics impact player experiences is vital to gaining a thorough understanding of how games affect them.

When employing the MDA framework's definition, dynamics encompass nearly every aspect of a game. 
As elucidated by one of the creators of MDA in an essay \cite{leblanc2006tools}, examining a game in terms of its dynamics entails asking the question, ``What happens when the game is played?'' 
Answering this question holistically for the entire domain of board games becomes overwhelming due to the multitude of events that occur within a single gameplay, not to mention the vast range of possible plays across all conceivable games.

To bridge the knowledge gap regarding dynamics, this study presents a comprehensive vocabulary specific to board game dynamics. 
This vocabulary was developed through the collaboration of a focus group consisting of experienced and novice players and was subsequently validated through a widely distributed survey. 
The outcome of this process includes a collection of dynamic names and a taxonomic structure to organize the vocabulary. 
It is important to note that our focus is solely on board games due to limitations in space and time. 
Defining the vocabulary for board games alone proved to be an arduous endeavor, and it is our hope that this initial effort will inspire other researchers to create a more exhaustive vocabulary.

The remainder of this paper is organized as follows: Section 2 elucidates the MDA framework; Section 3 outlines the methodology employed; Section 4 provides illustrations and analyses of the results; Section 5 discusses related works; Section 6 presents avenues for future research; and Section 7 concludes this study.

\section{the MDA framework}

The MDA framework stands for Mechanics, Dynamics, and Aesthetics. It was introduced by Robin Hunicke, Marc LeBlanc, and Robert Zubek in their paper ``MDA: A Formal Approach to Game Design and Game Research.''\cite{Hunicke2004} 
The framework provides a structured approach to understanding games by dissecting them into three interconnected components.

\begin{itemize}
    \item \textbf{Mechanics} refer to the specific rules, actions, and systems within a game. They encompass the tangible elements that players interact with, such as controls, objectives, resources, and the overall set of rules. Mechanics define the gameplay and serve as the building blocks of a game. They can include movement, combat, puzzles, character abilities, and more.
    \item \textbf{Dynamics} emerge from the interaction of mechanics within a game. They represent the behaviors, patterns, and responses that arise as players engage with the mechanics. Dynamics encompass the player's experience of playing the game, including the strategies, challenges, and outcomes that occur during gameplay. These can involve decision-making, emergent gameplay, competition, cooperation, and the overall flow of the game.
    \item \textbf{Aesthetics} capture the emotional responses, moods, and overall player experience evoked by the game. They encompass the subjective qualities that make a game enjoyable, such as immersion, fantasy, narrative, audiovisual elements, and overall atmosphere. Aesthetics focus on the player's emotional engagement and the intended experience that the game designers aim to create.
\end{itemize}

The MDA framework emphasizes the interplay and iterative nature of game design. Mechanics drive the dynamics, which, in turn, evoke specific aesthetic experiences. However, aesthetics can also influence the design of mechanics, as designers may tweak or introduce new mechanics to enhance the desired emotional experience. This iterative process allows designers to fine-tune the game's mechanics and dynamics until they align with the intended player experience.

\section{Methodology}


Our work brought together quantitative and qualitative research to provide an initial understanding of the dynamics of board games. 
This was done by first using a focus group composed of game designers to propose names and concepts of dynamics, and then using a survey more widely applied to validate the results of the focus group and provide more suggestions. 
After the initial set of concepts was evaluated through the survey results, it was further pruned when needed, and after such filtering they were used to compose the vocabulary of dynamics.

\subsection{The focus group}

A focus group is a method to generate data based on individual experiences and the discussion of such individuals on those experiences. 
These individuals should have common expertise in the topic to be addressed in the experiment. 
Even better is if the participants have a good amount of knowledge and experience on the subject. 
According to \cite{jenny_methodologyfocusgroup_1994} the most important data of a focus group is provided by the interaction and discussion among its members.
However, a focus group has to remain focused on the topic to be addressed and thus needs to be directed correctly to provide better quality data for the research. \cite{liamputtong_focusgroup_2011,rabiee_focus-group_2004,jenny_methodologyfocusgroup_1994}

The selection of individuals for this focus group should, of course, be composed of the target users of this methodology, i.e., in our case, it must include board game designers. 
Also, it intends to collect data on dynamics, which happens during the play of the game. 
Thus long-time players of board games should also be able to provide valuable insight and are included in this selection. 

The designers and testers of the collective {\textit{Casa do Goblin}}\footnote{{\url{www.casadogoblin.com}}} collective agreed to be our focus group. 
The ten participants were made up of 5 men and 5 women, 23 to 40 years old.

Although experts in board games, the participants had different notions of dynamics or did not know at all the specific definition provided by the MDA model.
As such, it was needed to thoroughly explain the dynamics concept of the MDA to them. 
After that, the group brainstormed on names or terms of dynamics that they believe pertain to board games. 
Moving on, they discussed what was proposed in the brainstorming to filter out unfitting concepts and to further develop those that required extra attention.

The discussion was directed towards generating words for dynamics, that is, to explain concepts they believe to be dynamics into simple words. 
To provide an anchor from which to start, the mechanics featured in \cite{kritz_buildingOntology} were fixed on a board in everyone's sights. 
The purpose of the mechanics was to be used as the basis for thinking of dynamics they found in games, that is, looking at the mechanics and thinking about which dynamics emerge from each of them.

\subsection{Survey on dynamics}

At the beginning of the survey, there was a short explanation of what a dynamic board game is according to the MDA. 
This explanation was provided to ensure that all subjects shared the same understanding of the concepts.
Additionally, for statistical purposes and further analysis, there were questions about the subjects' relationship to the board game world (designer, producer, hobbyist) and how long they had known and played board games. 
This distinction was based on the MDA proposal, which suggested that players and designers have different perspectives on the game, leading to potential differences in their opinions about dynamics.

The survey was composed of questions based on the results of the focus group. For each term created, the survey included traditional 5-point Likert scale questions. The subjects surveyed were asked to indicate their level of agreement or disagreement with each term as a dynamic that actually occurs in board games. The inclusion of a neutral option between agreement and disagreement allowed answers such as ``not knowing'' or ``not understanding.'' To verify the information of the respondents, the survey required details on their relationship to board games and the frequency of their gameplay \cite{devellis2016scale}.

Furthermore, participants had the option to suggest new dynamics that they felt were missing and provide an email address to receive further information on the research if they were interested.

The survey, available in English and Portuguese versions, was shared through various channels, including Facebook and WhatsApp groups focused on board games, as well as the boardgamegeek forum. This multilingual approach aimed to gather a greater number of responses and ensure a diverse range of opinions. We used Google Forms to create and administer the survey. The data collection phase concluded after one week of survey release. The surveys used can be found in our GitHub repository\footnote{{https://github.com/LUDES-PESC/DynamicsVocabulary}}.


\section{Results}

This section evaluates the results of the experiments as well as construes the vocabulary of dynamics.

\subsection{Focus Group}
Results from the focus group were used to create a basis of knowledge on the dynamics of board games. This knowledge was represented in the form of 57 ideas.
The analysis of the ideas collected and the compilation of them into a more compact list was done shortly after, which was important before these ideas could be used in subsequent tasks. 

Primarily, the group proposed ideas of dynamics. 
It developed several concepts, which were then discussed and checked whether they were dynamics or not. 
Many concepts were mentioned by more than one person, even in different terms. 
This means that the group was well aligned in how they understood the dynamics. 
Other concepts caused great dissension as to whether they were dynamics or not. 
Further studies should be made to evaluate this; thus, they were excluded from this work. 
The group could not associate some ideas, which seemed to be dynamics, with a proper mechanic. 
Without such a connection, they were not included in this ontology. 
The discussion of the concepts proved to be the most valuable. Some of them did not appear to be dynamics a first glance. But in another perspective or situation, they were dynamics.

Apart from the dynamic concepts created, the discussion featured their relationships to mechanics and aesthetics. As a group, the participants related many dynamics to the other concepts when explaining them or trying to understand them. Sometimes dynamics are also related to each other. Establishing that some are incompatible with each other, and generalizations between them.

Provided the initial terms and concepts, the group then reviewed the terms generated and discussed what they mean, how they present themselves in games, and whether it is or not a dynamic. 
As needed, small adjustments were made, and the terms became 57 concepts to be used in the vocabulary.

In order to streamline the survey and avoid a lengthy questionnaire with obligatory questions, a trimming process was implemented. 
Since 57 concepts resulted in 57 questions, it was necessary to reduce the number of concepts. 
Redundant concepts, those contained within one another, were removed, leaving only the more generic ones. Furthermore, similar concepts that could be abstracted into a new concept not present in the list were eliminated, favoring the more general concept. 
This adjustment was aimed at increasing the likelihood of receiving an adequate number of responses. \cite{malhotra2012pesquisaMarketing}

Finally, a list of 40 dynamics was achieved. The complete list can be found in our GitHub repository \footnote{{https://github.com/LUDES-PESC/DynamicsVocabulary}}

\subsection{Curious cases}

Some concepts that arose during the focus group discussions were very interesting although they did not reach the final list. This was because they were very complicated cases. But the fact that they were brought to discussion is by itself a great contribution. 

One was Chaos, meaning a condition of disorder in the
game, that is, chaos made by the players. It was very
surprising to most, and through discussion, they could not
agree whether it is or not a dynamic. The most interesting
factor in this discussion, however, is that if it is a dynamic,
other possible conditions, like funny, tense, and ordered,
would also be dynamics. On the other hand, it is also
possible that these concepts are actually part of Aesthetics,
as defined in the MDA\cite{Hunicke2004} as they could be associated with emotional states. 

Another case is the change of rules during the game.
This situation can occur in the game in some situations,
more notably when players notice confusion or a wrong
interpretation of the rules. It could also happen when players
find something broken while creating games, and it is
common in children’s play. The first case always raises the question ``do we continue to play wrong to avoid changing
the game as it is now, or from now on we play correctly?''.
Hence the possibility of being a dynamic, since it comes
from the decisions the players make. However, since it is
not part of the gameplay itself, it can be considered a meta-game event.

The focus group also detected another common situation
in board games that can be considered meta-game action:
a player needing or wanting to undo his last move or play,
including the case that it was impossible to be made with
the correct rules. 

All of those situations are left unanswered in this work.
They push the boundaries of our scope of
preliminary work, requiring more study and thought, and
were left for future development.

\subsection{Survey}

The responses to the survey were collected for one week, from 12/13/2018 to 12/20/2018. Featuring a total of 196 answers, which were divided into two groups: consumers with 174 responses, and industry totaling 22 answers. Then the average value for each question in each group was calculated. These values were omitted due to space constraints.

Looking for the concepts which were mostly agreed upon by the subjects, that is, with an average equal to or higher than 3, becomes possible to observe an interesting phenomenon. The industry group had 32 concepts at this level and the consumer group had 29. It is important to note that 29 of the consumers are present in the 32 of the industry. That is, consumers disagree only on three concepts from the industry, which lead to the belief that their opinions are similar or equal. This was a surprise to us because of our initial belief that they would have different opinions. Analyzing the concepts with even more average values, those with a value greater than or equal to 4, we have almost the same situation. With 14 concepts for industry and 9 for consumers, we note a bigger difference in the number of higher values. Also, of the consumers' 9, only one of them is not contained in the 14 from industry. Although it still poses good evidence that the group's opinion is in accordance, it also points to some doubt in this assertion. To enlighten this doubt, this work uses the ANOVA \cite{malhotra2012pesquisaMarketing} test to state if the groups have significantly different results.

\subsection{Statistical analysis of survey answers}

The purpose of ANOVA is to test the hypothesis that different groups of samples for the same factor have a statistically significant difference in the results. This will give the information needed to establish the correct average values for the concepts. As with any model, we need to be inside its constraints to be able to use it. One-way ANOVA has only one crucial constraint, the variances of the residues must be equal. Testing it, with a proper method, is very simple and provides a probabilistic answer. Using a standard significance level of $\alpha = 0.05$ provides a good perspective on the similarity of the variances.

Using the equal-variance test in our data provided us with a 95,8\% certainty that the variances are equal, according to the multiple comparisons tests. This test takes into account the intervals of each variance of the sample. When they do not overlap, there is a significant difference in the variances. Levene's test provided a P-Value of 0,971, meaning that there is a 97,1\% chance of the variances being equal. Using two different tests and obtaining a positive result on both of them, we are now assured that we can use one-way ANOVA. 



The raw results of the test are shown in Table \ref{tab:AnalVar}. These results have many parts: the main result of the model calculations and the model adequacy measurements also presents some information regarding the data. The resulting calculations of the model under the Analysis of Variance show us the value of the sums of squares and the sum of means squares. With them, the model evaluates the F-Value, of $0,83$, and the P-Value, of $0,366$. Both of them do not reject the null hypothesis of the model, as the p-value is way over our alpha. 

Adequacy is shown as a summary of the model in Table \ref{tab:ModelSum}, S is the distance between the data and the model, and R2 is the percentage of variability in the results presented by the model.

The results show that the model did not refute the hypothesis that the averages are equal. Thus, allowing us to look at the data of both groups together. That is, to average all the answers without concern for the group it came from.

\begin{table}[h]
    \centering
    \caption{Analysis of Variance}
    \begin{tabular}{c | c c c c c}
     Source & DF & AdjSS & Adj MS & F-Value & P-Value\\
     \hline
     Factor & 1 & 0,4147 & 0,4147 & 0,83 & 0,366 \\
     Error & 78 & 39,1889 & 0,5024 \\
     Total & 79 & 39,6036
     
    \end{tabular}
    
    \label{tab:AnalVar}
\end{table}

\begin{table}[h]
    \centering
     \caption{Model Summary}
    \begin{tabular}{c c c c}
     S & R-sq & R-sq(adj) & R-sq(pred) \\
     \hline
    0,708817 & 1,05\% & 0,00\% & 0,00\%
     
    \end{tabular}
   
    \label{tab:ModelSum}
\end{table}


\subsection{Resulting Vocabulary}

After consideration of the analysis result, we concluded that averaging both our populations as equals was the best course of action.

With this, a final average level for each concept was calculated. The goal was to extract from the original list the core game dynamics experienced in boardgames, making the assumption that they are represented by concepts mostly agreed upon on the survey. 

To put concepts together in this effort they also needed to be synthesized in a term, since vocabularies are made of terms that are linked to a meaning. With that in mind, here is the vocabulary created, composed of 29 of the 40 concepts featured in the survey:

\begin{multicols}{2}
    
\begin{itemize}
    \item Sacrifice: Sacrifice a piece or position for greater gains
    \item Indirect Effect: Execute something when wanting a consequence of its events
    \item Acquire information: Use action to discover 
    \item Reduce Options: Using an action to reduce other players' options
    \item Resource Extinction: Reduce the source of a limited resource or make it useless
    \item Deduction: Use open information to discover hidden information
    \item Game state Change: Use action to purposely provoke a change in the game
    \item Combo: Chain automatic effects of the game
    \item Blocking: Block another player's action, strategy, progress
    \item One versus all: When one player attacks all the others simultaneously 
    \item All versus one: All players of the game, but one, join forces to defeat the other player
    \item Alliance: When players join forces to achieve mutual benefit
    \item Forceful interpretation: Use a particular point of view to create better benefits for himself
    \item Self-objective: Pursue a self-appointed objective other than the game's objective
    \item Play safe: Not taking risks, playing only on certainty
    \item Risk play: Accept greater risks seeking greater rewards
    \item Survival: Play only with to avoid elimination
    \item Camping: Stick with a position or action for a lot of time
    \item Protectionism: Protect a specific position or pieces
    \item Action planning: Play accordingly to your next actions, and plan a series of actions.
    \item Rush the game: Accelerate the end of the game
    \item Flexible strategy: Change strategy because of the game state
    \item Reject objectives: Intentionally not achieving a game's objective to attain some advantage.
    \item Intimidate: Use a stronger position to force another player to play as you want
    \item Distraction: Use an action to change other players' attention from your real intention or objective
    \item Small talk: Talking all the time to distract other players
    \item Count resources: Use previous knowledge of the available resources to count them and achieve an advantage
    \item Bluffing: Relay false information to manipulate other players' actions
    \item Convince: Convince other players
\end{itemize}
\end{multicols}

We structured these concepts of dynamics by grouping the concepts by similarities which provides a taxonomic structure to dynamics. To define these categories we used both the authors' experience and the discussion made during the focus group. This process resulted in the following categories:

\begin{itemize}
    \item Action based: generalizes the dynamics which are based on the agency of the players.
    \item Intention of use: generalizes the dynamics which arise from a specific intention when using an action.
    \item Meta-game: generalizes the dynamics that happen outside of the game space but still inside the magic circle.
    \item Behaviour: generalizes the dynamics that represent a particular behavior a player can adopt.
    \item Playing patterns: generalizes the dynamics which amount to how a player plays the game for some time or even the whole game.
    \item Strategy choices: generalizes the dynamics that evaluate a specific play or principle used momentarily during the game.
\end{itemize}

\begin{figure}
    \centering
    \includegraphics[width=\textwidth]{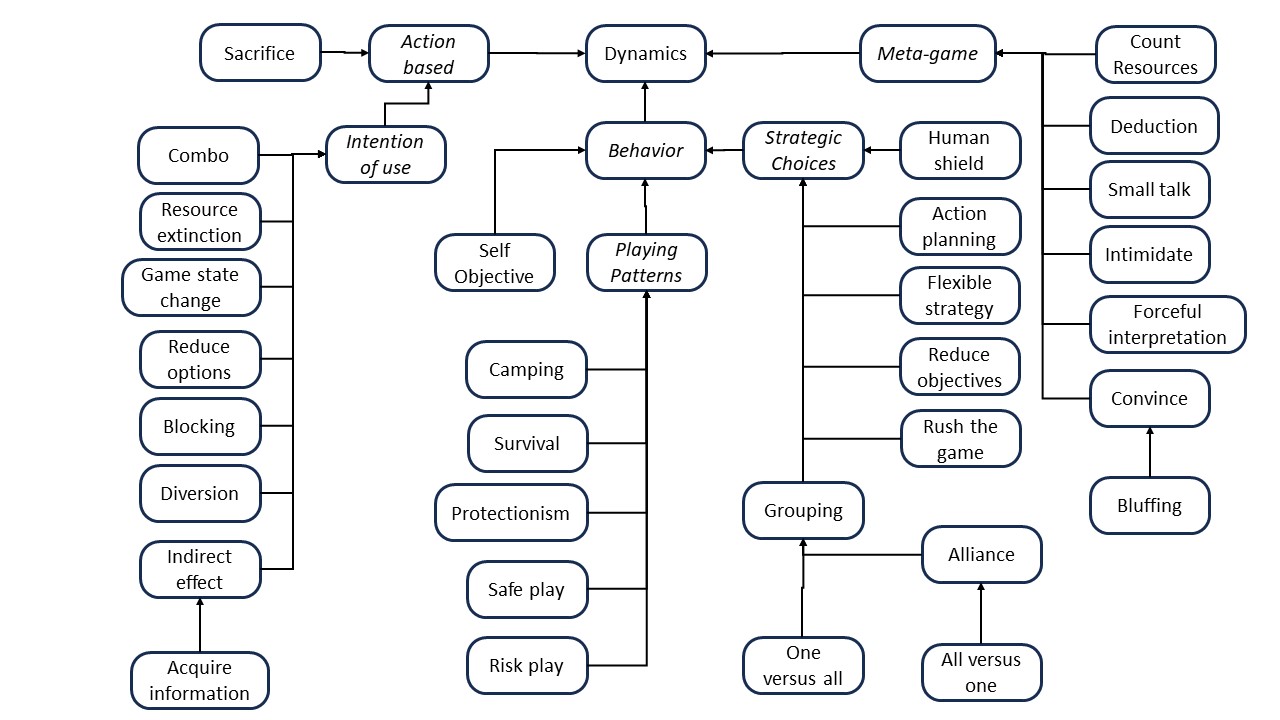}
    \caption{Dynamics vocabulary}
    \label{fig:dynVoc}
\end{figure}

This structure is illustrated in Figure ~\ref{fig:dynVoc}

\section{Related Work}

Attempts to classify and understand game concepts are abundant in the literature; however, these efforts are directed at concepts other than dynamics. 
One common approach is the creation of ontologies, and there are many attempts to create such an artifact. 
The most known of these attempts is the Game Ontology Project (GOP) \cite{wiki:gop}: ``a framework for describing, analyzing and studying games'' \cite{Zagal:2008:GOP}. 
It provides a structure to study game elements based on four top-level elements:  interface,  rules, entity manipulation, and goals. 
It is a collaborative work open to contributions, although at least since 2019, its development has stalled. 
Roman,  Sandu,  and  Buraga  constructed  an  ontology  for  role-playing games (RPG) in a work inspired by GOP \cite{roman2011owl}. 
A slightly more comprehensive ontology was created in the realm of RPGs \cite{dhuric2015specific}. 
Although  using  a  specific  game,  \cite{manaworld}, as the source of concepts, the authors claim that the resulting ontology is applicable to massively multiplayer online role-playing games (MMORPG). These are some examples, but there are many other game ontologies, such as: \cite{innov:gop:mda,300gm,leon_z._ontology_2010,parkkila_ontology_2017,sacco_game_2017},
\cite{kritz_buildingOntology}.

Another form of categorization of elements is catalogs. An example close to our work is \cite{engelstein2019building}, which is also applied to boardgames, but comprises only mechanics. 
Bjork and Holopainen created a catalog of game design patterns\cite{bjork_patterns_2005}. 
The Art of Game Design provides both a catalog of lenses that aim to evaluate games and a theoretical framework that divides games into four elements: Mechanics, Aesthetics, Technology, and Story \cite{schell2014art}. Samarasinghe et al. used a data-driven approach to create and analyze a data set of mechanics and design patterns for boardgames \cite{samarasinghe2021data}.

Finally, we have frameworks that separate games into elements that are then further explained or defined. An example is the theoretical framework used in this work, the MDA framework \cite{Hunicke2004}. However, there are many other frameworks used in the literature: \cite{schell2014art,salen2004rules,jarvinen2009games,juul2010game}.

The only work focusing mainly on game dynamics was a continuation of the MDA framework made by LeBlanc in \cite{leblanc2006tools}. However, there are a number of works focusing on other aspects of games that mention dynamics \cite{schell2014art,salen2004rules} although they usually group them with mechanics or just do not develop them in any way. On the other hand, some of these catalogs define mechanics that are, in fact, dynamics. Egelstein and Shalev include deduction and induction in their second edition of \cite{engelstein2019building}.

\section{Conclusion}

This work aimed to develop the game dynamic domain. We did so by creating and presenting a vocabulary of dynamics present in boardgames. This vocabulary is by no means comprehensive, and even more, it is only preliminary. However, it provides a first step in the development of the game dynamics domain.

The result of this work is useful to game designers in that they can now more effectively discuss dynamics present in their games. For researchers, this contribution is two-fold: It can be incorporated into existing and future game frameworks to provide a more complete representation of games, and it enhances the analytical possibilities of any game evaluation by allowing researchers to name parts of the game in a consistent form.

However, the greatest advantage of this work is the ability to support future research. Such as increasing the vocabulary and aiming for a more comprehensive one, creating new frameworks centered on dynamics rather than mechanics. Strengthening player-centered research, as dynamics are closer to players than mechanics. Many other lines of research are viable, some of which are still unknown to us. Nevertheless, we have one certainty: game dynamic is a domain filled with potential to enhance game studies and should not be neglected anymore.

\bibliographystyle{sbc}
\bibliography{main}

\end{document}